\documentclass{WileyMSP-template}
\usepackage{amsmath,amssymb}
\usepackage{color}
\usepackage{textgreek}

\begin{document}

\pagestyle{fancy}
\rhead{\includegraphics[width=2.5cm]{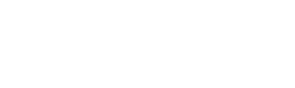}}

\title{Tracing boron diffusion into a textured silicon solar cell using electron beam induced current in a scanning transmission electron microscope}
% by combining boron diffusion simulation with experimental and simulated scanning transmission electron beam induced current

\maketitle

% Author: Please give full first and last names for authors and include * after the name of all corresponding authors

\author{Tobias Meyer $^\dagger$*}
\author{David A. Ehrlich $^\dagger$}
\author{Peter Pichler}
\author{K. L. Skrollan Detzler}
\author{Christoph Flathmann}
\author{Valeriya Titova}
\author{Tim Böckendorf}
\author{Hartmut Bracht}
\author{Jan Schmidt}
\author{Michael Seibt} 

% Dedication

\dedication{}

% Affiliations: Please provide adacemic titles (Prof. or Dr.) for all authors where applicable, and include an institutional email address for all corresponding authors
\begin{affiliations}
T. Meyer, D.A. Ehrlich, K.L.S. Detzler, C. Flathmann, M. Seibt\\
4th Institute of Physics –Solids and Nanostructures, Georg-August-University Goettingen, Friedrich-Hund-Platz 1, Göttingen 37077, Germany\\
Email Address: tmeyer@uni-goettingen.de

T. Meyer\\
Institute of Materials Physics, Georg-August-University Goettingen, Friedrich-Hund-Platz 1, Göttingen 37077, Germany\\

P. Pichler\\
Fraunhofer Institute for Integrated Systems and Device Technology IISB, Schottkystrasse 10, 91058 Erlangen, Germany\\
Universit\"at Erlangen-N\"urnberg, Lehrstuhl für Elektronische Bauelemente
Cauerstrasse 6, 91058 Erlangen, Germany\\

C. Flathmann\\
Research Center Future Energy Materials and Systems, Ruhr University Bochum, Universitätsstr. 150, Bochum 44801, Germany\\
Faculty of Physics and Astronomy, Ruhr University Bochum, Universitätsstr. 150, Bochum 44801, Germany\\

V. Titova, J. Schmidt\\
Institute for Solar Energy Research Hamelin (ISFH), Am Ohrberg 1, 31860 Emmerthal, Germany\\
Institute of Solid-State Physics, Leibniz Universität Hannover, Appelstrasse 2, 30167 Hannover, Germany\\

T. Böckendorf, H. Bracht\\
Institute of Materials Physics, University of Münster, Wilhelm-Klemm-Str. 10, 48149 Münster, Germany\\

{$^\dagger$ These authors contributed equally to this work.}
\end{affiliations}

% Keywords: Please provide a minimum of three and a maximum of seven keywords, separated by commas


% Abstract should be written in the present tense and impersonal style (i.e., avoid we), and be at most 200 words long
\begin{abstract}
The light absorption of [001] grown single-crystalline silicon wafers can be enhanced by chemical etching, e.g., with potassium hydroxide, resulting in a pyramid-like surface texture. Alongside advantageous photon harvesting in solar cells, the surface roughness leads to drawbacks when measuring diffusion behaviour of dopants in the heterogeneous structure. In this paper, we employ experimental and simulated scanning transmission electron beam induced current in combination with simulation of boron diffusion in a self-consistent framework to trace the dopant distribution underneath the pyramid-like surface texture.
In order to account for surface recombination, an effective model projecting the system along the electron beam propagation direction is used in the EBIC simulation enabling a comparison to entire two-dimensional experimental maps. We find a good agreement between simulated and experimental data and thoroughly discuss how EBIC can be used in future experiments to quantify weak electric fields.
\end{abstract}

% Text: Please use section headings and subheadings as specified below. For communications, all section headings apart from Experimental Section should be removed
% Please make the first reference to a display item bold: \textbf{Figure 1}
% Do not abbreviate Figure, Equation, etc.; display items are always singular, i.e., Figure 1 and 2.
% Equations are always singular, i.e., Equation 1 and 2, and should be inserted using the {equation} environment, not as graphics
% Please do not use footnotes in the text, additional information can be added to the Reference list.

\section{Introduction}

The decreasing size and mounting design complexity of modern optoelectronic devices imposes great challenges to fabrication processes as well as subsequent characterization techniques.
Prominent examples are lower-dimensional systems like quantum dots \cite{wu2025situ}, rods \cite{prodanov2025highly}, and wells \cite{liu2023zwitterions}, which demand for methods offering high spatial resolution and sensitivity to electronic properties.
Furthermore, strongly textured devices can involve spatially heterogeneous properties that are not fully captured by macroscopic characterization.
For instance, pyramid-like surface textures are used to increase photon absorption in silicon solar cells \cite{Singh2019}. However, profiling the concentration of dopants incorporated in subsequent diffusion processes
is strongly impeded.
In fact, methods with low lateral resolution like electrochemical capacitance-voltage \cite{peiner1995doping}, plasma time-of-flight mass spectrometry \cite{spende2020plasma}, or secondary ion mass spectrometry (SIMS) \cite{mcphail2006applications} are not applicable in the presence of strong textures. 
Furthermore, techniques like atom probe tomography \cite{gault2021atom,rybak2025atomic} or nano SIMS \cite{li2020nanosims,bernzen2024potassium} that do achieve high lateral resolution but rely on mass spectrometry detect dopant atoms irrespective of their ionization state which matters most for electronic applications.
On the other hand, while offering hight spatial resolution and sensitivity to active dopants, achieving consistent calibration in scanning spreading resistance measurements (SSRM) \cite{eyben2002scanning} and scanning capacitance measurements \cite{williams1999two} remains challenging due to tip wear.
Alternatively, electron beam induced current (EBIC) is used to trace dopant distribution in textured devices \cite{Ma2014,zhou2020understanding} and recently the method -- traditionally conducted in a scanning electron microscope (SEM) \cite{leamy1982charge,hackett1972electron,jastrzebski1975application,berz1976theory,donolato1983evaluation,luke1985quantification,ong1994direct,kittler1996ebic,sekiguchi2002characterization,saring2013electronic} -- was transferred to the scanning transmission electron microscope (STEM) \cite{Cabanal2006,peretzki2018implementation,Meyer2019,meyer2020structural,duchamp2020stem,Conlan2021,schneider2025stem} offering atomic scale spatial resolution.
Due to the strong effect of surface recombination, however, a straightforward quantitative interpretation of EBIC signals is difficult and several effective models accounting for bulk and surface recombination in different geometries were reported. Importantly, a quantitative model applicable to the STEM-EBIC geometry, i.e., thin, electron-transparent samples, was recently demonstrated in \cite{meyer2026quantitativemodelsexcesscarrier}.

Complementary to the mentioned experimental techniques, the diffusion of dopants \cite{Pichler2012} -- in particular boron \cite{Antoniadis1978,Mirabella2013,bracht2007self,bracht2007self2} -- in silicon was studied extensively on planar samples to determine corresponding diffusion parameters which can be used subsequently as input parameters for the simulation of heterogeneous structures. 
Using this approach, an enhanced boron diffusion in the case of an oxidizing ambience was found in \cite{Antoniadis1978}.

In this work, we combine STEM-EBIC with diffusion simulations to trace the boron distribution in a textured silicon solar cell. A self-consistent framework is established by comparing experimental STEM-EBIC results to simulated signals obtained with the finite element method (FEM) and using simulated boron distributions as model input. Employing an effective recombination model including bulk and surface recombination as suggested in \cite{meyer2026quantitativemodelsexcesscarrier}, we demonstrate for the first time that entire STEM-EBIC maps can be modeled numerically enabling unprecedented comparisons between experiments and simulations in heterogeneous systems.
Lastly, strategies how STEM-EBIC can be employed in the future to study recombination mechanism in space charge regions as well as to measure weak electric fields quantitatively are discussed.

\section{Sample Preparation and Instrumentation}

\begin{figure}[t]
\includegraphics[width=.99\textwidth]{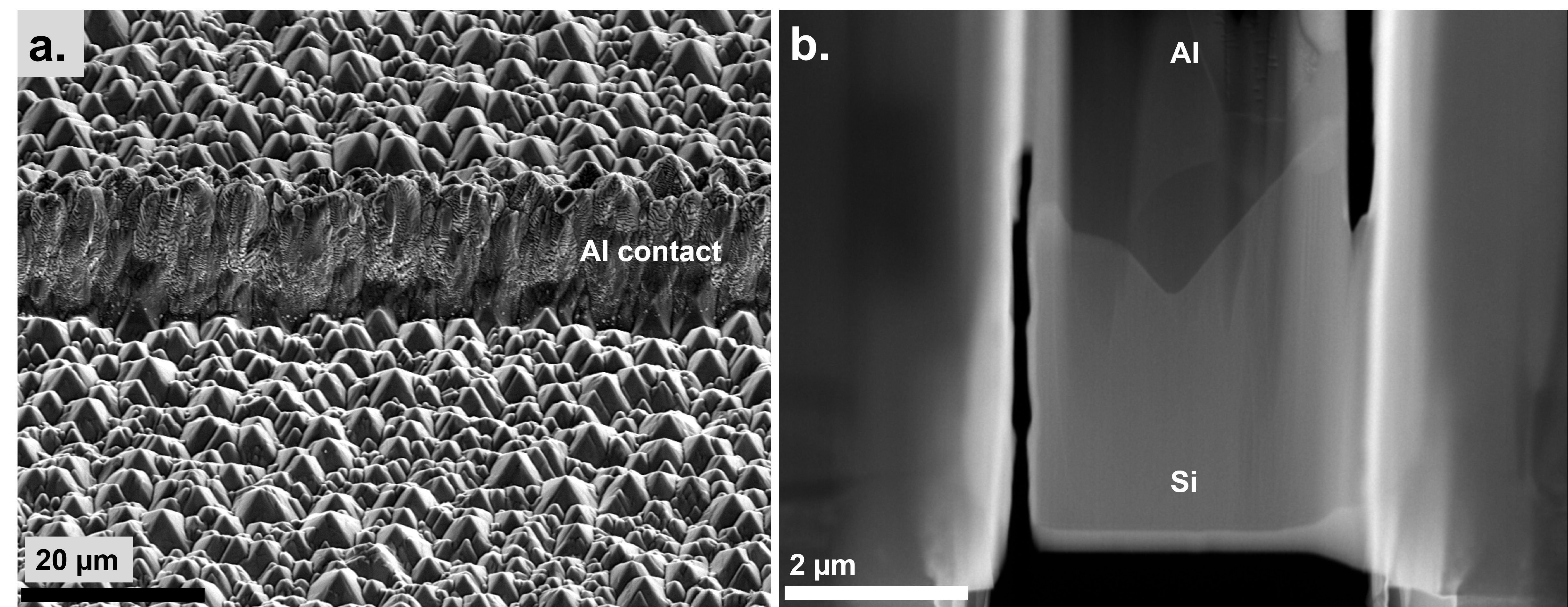}%
\caption{(a) SEM overview of the pyramid texture of the solar cell including an aluminium front contact. The surface was inclined by 45 degrees during imaging. (b) ADF-STEM overview of the extracted lamella in [110] zone axis including a trench between two pyramids in the central region.
Vertical cuts have been applied to prevent short circuits across the investigated area.}
\label{fig:sample_prep}
\end{figure}

A detailed description of the preparation process of the silicon solar cell can be found in~\cite{cell}. Briefly, the main steps related to the boron diffusion process conducted in a Tempress Systems TS-81004 furnace
%being relevant for the boron diffusion simulation later on 
are summarized as follows:
\begin{enumerate}
    \item Heat-up: The furnace temperature is increased from 700\,$^\circ$C to 941\,$^\circ$C using a ramping speed of 10\,$^\circ$C/min in an inert ambience.
    %and subsequently stabilized for further 15\,min at a 10\,Pa below ambient pressure and an N$_2$ gas flow of 15\,slm.
    \item Deposition: Keeping the temperature %and pressure 
    in the furnace stabilized, 
    %400\,sccm 
    N$_2$ is lead through a liquid BBr$_3$ bubbler held at 20\,$^\circ$C leading to a B$_2$O$_3$ flow which is subsequently mixed with N$_2$ and O$_2$.
    %a gas flow of 15\,slm N$_2$ and 160\,sccm O$_2$ for 25\,min.
    \item Drive-in: The flow through the bubbler is switched off while all other parameters remain unchanged for further 20\,min.
    \item Post-oxidation: The N$_2$ and O$_2$ gas flows are reduced and the temperature is held constant for further 60\,min.
    %changed to 5\,slm N$_2$ plus 10\,slm O$_2$ and the pressure reduced to 25\,Pa below ambience for 60\,min.
    \item Cool-down: %Using a gas flow of 15\,slm N$_2$, 
    In an inert ambience, the temperature is decreased by -10\,$^\circ$C/min and the sample carrier ultimately retracted at 805\,$^\circ$C.
\end{enumerate}
% \begin{enumerate}
%     \item Heat-up: The furnace temperature is increased from 700\,$^\circ$C to 941\,$^\circ$C using a ramping speed of 10\,$^\circ$C/min and subsequently stabilized for further 15\,min at a 10\,Pa below ambient pressure and an N$_2$ gas flow of 15\,slm.
%     \item Deposition: Keeping the temperature and pressure in the furnace at the stabilized values, 400\,sccm N$_2$ are lead through a liquid BBr$_3$ bubbler held at 20\,$^\circ$C and subsequently mixed with a gas flow of 15\,slm N$_2$ and 160\,sccm O$_2$ for 25\,min.
%     \item Drive-in: The flow through the bubbler is switched off while all other parameters remain unchanged for further 20\,min.
%     \item Post-oxidation: The gas flow is changed to 5\,slm N$_2$ plus 10\,slm O$_2$ and the pressure reduced to 25\,Pa below ambience for 60\,min.
%     \item Cool-down: Using a gas flow of 15\,slm N$_2$, the temperature is decreased by -10\,$^\circ$C/min and the sample carrier ultimately retracted at 805\,$^\circ$C.
% \end{enumerate}
In addition to the textured wafer, a planar reference sample was processed simultaneously and investigated by four point probe (4PP) measurements subsequently to determine the sheet resistance of the resulting boron doped layer in the homogeneous case. 

An SEM overview of the textured surface including an aluminium front contact is shown in Figure \ref{fig:sample_prep}a. An electron transparent lamella was extracted from the aluminium covered area using an FEI Nova NanoLab focused ion beam (FIB). The acceleration voltage for the final thinning step was decreased from 30\,kV to 5\,kV in order to minimize beam damage at the surfaces. 
An annular dark-field (ADF-) STEM overview of the resulting lamella is shown in Figure \ref{fig:sample_prep}b. The vertical cuts have been applied during FIB preparation to prevent short circuits across the electrical junction due to redeposited material.
For electrical contacting, a DENSsolutions Lightning D7+ holder was used. All STEM experiments were performed inside an FEI Titan 80-300 E-TEM operated at 300\,kV and using a convergence semi-angle of 10\,mrad.
The beam current was set to 42\,pA and a collection semi-angle of 39\,mrad was used for electron energy loss spectroscopy (EELS) employing a Gatan Quantum 965 ER. Thickness values were determined via EELS using the log-ratio method and an inelastic mean free path of 182.4\,nm for silicon following the parametrization in \cite{iakoubovskii2008thickness}.
The short circuit current was measured using a Stanford Research Systems SR570 current preamplifier.

For cross-sectional SSRM measurements, the sample was cleaved and a protective silicon wafer glued on top. Subsequently, the sample was polished using Al$_2$O$_3$ solutions with 15\,\textmu m, 5\,\textmu m, and 1\,\textmu m grain sizes, followed by polishing using diamond pastes with 1\,\textmu m and 0.25\,\textmu m grain sizes. 
A final polish was realized using a chemical-mechanical polishing solution (\textit{Köstrosol 3550}). The SSRM backside contact was connected using eutectic indium-gallium. SSRM was measured with a Park Systems XE-100 using Adama Innovations AD-40-AS cantilevers and a FEMTO DLPCA-200 current amplifier.

\section{Experimental Results}

\begin{figure}[t]
\includegraphics[width=.99\textwidth]{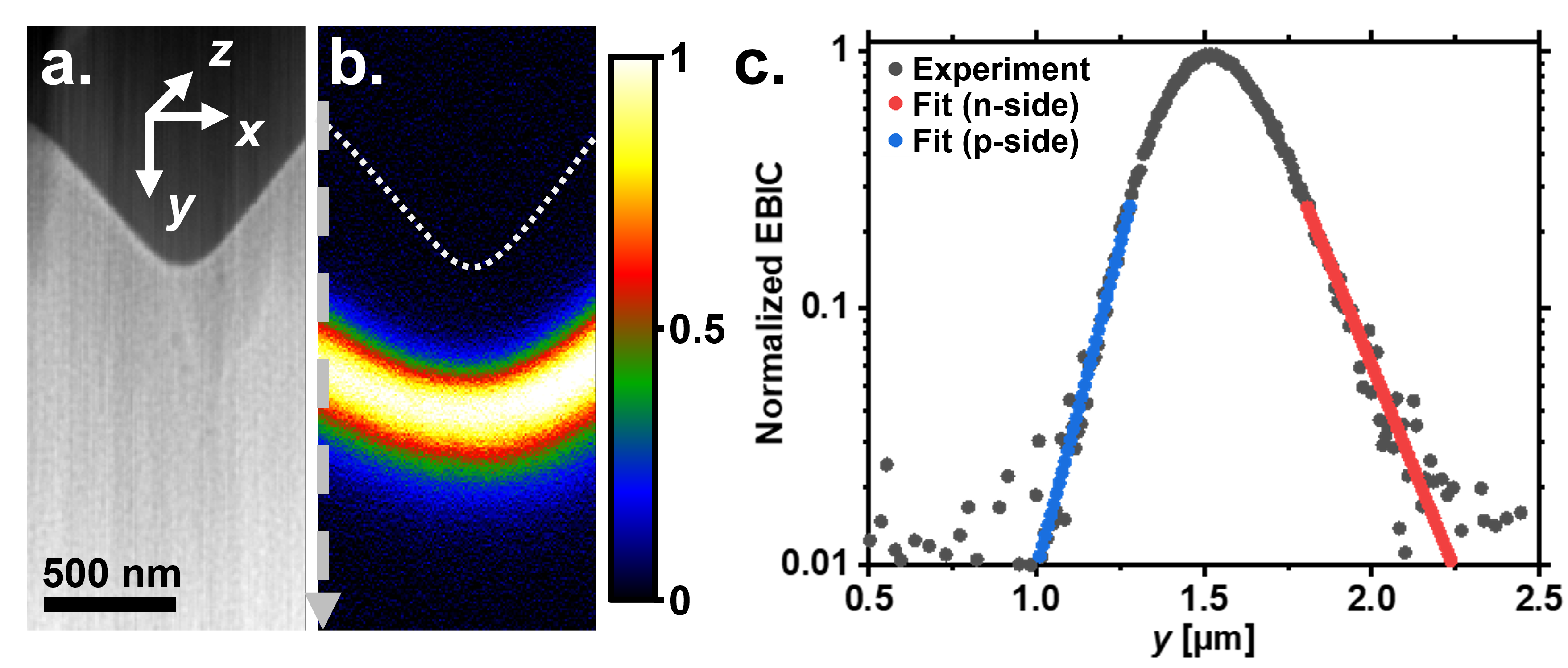}%
\caption{(a) ADF-STEM image of the central region of the lamella shown in Figure \ref{fig:sample_prep}b and (b) the corresponding EBIC map normalized to its maximum. The coordinate system used throughout the paper is indicated in (a). (c) Vertical EBIC profile along the dashed gray arrow in (b) as well as exponential fits of the tails on the n- and p-side. Only values below 0.25 have been considered during fitting.}
\label{fig:experiment}
\end{figure}

\begin{figure}%
\includegraphics[width=.99\textwidth]{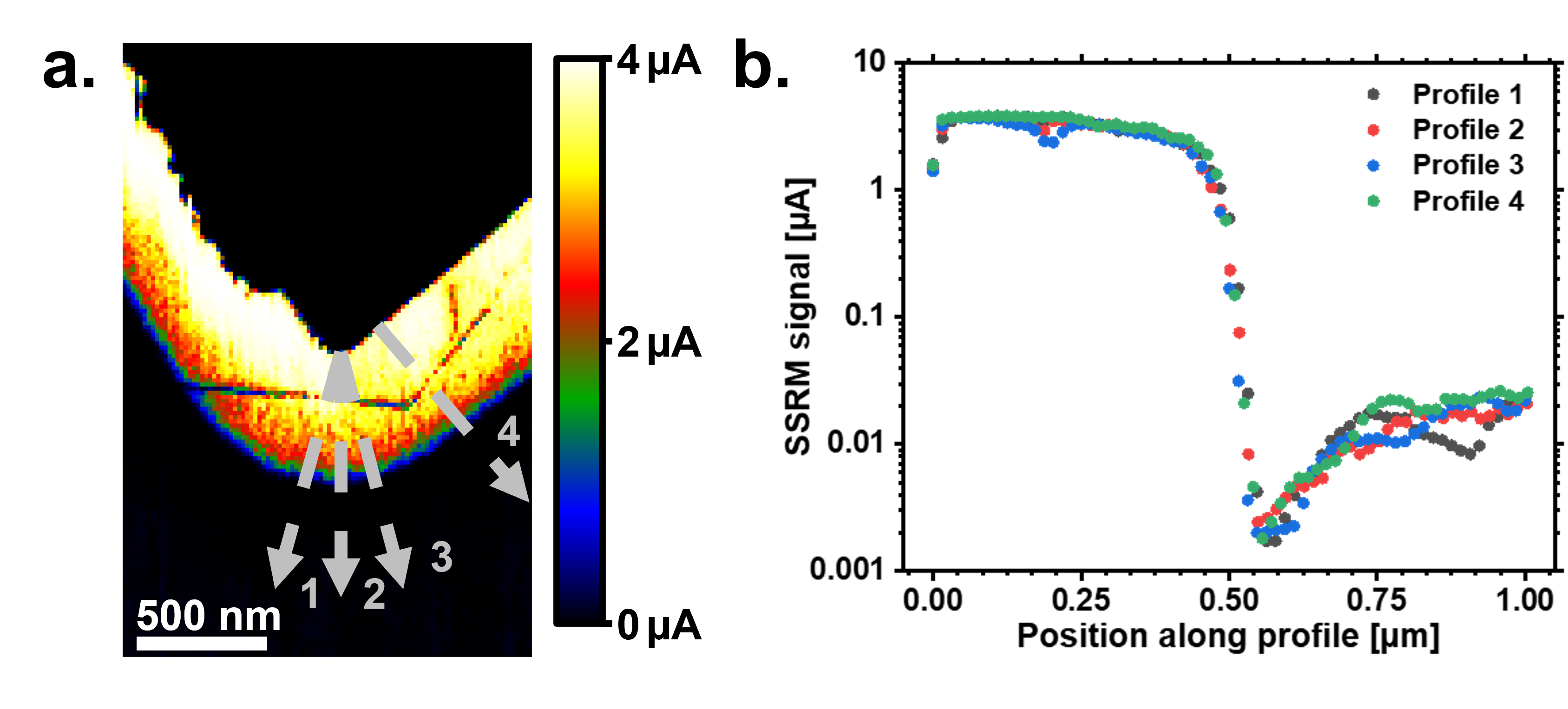}
\caption{(a) Cross-sectional SSRM current map (at 125 mV applied bias voltage) including a trench of the pyramid texture. (b) Line profiles taken along the arrows indicated in (a).
Please note that the shown data was collected at a different sample position than the results in Figure \ref{fig:experiment}.
High current values are obtained in the boron-doped region and local minima occur approximately at the junction.}
\label{fig:ssrm}
\end{figure}

The central part of the lamella shown in Figure \ref{fig:sample_prep}b including a trench between two pyramids was investigated simultaneously by ADF-STEM and EBIC yielding the maps presented in Figure \ref{fig:experiment}a and \ref{fig:experiment}b. The EBIC data is normalized to its maximum and a clear signal is found well below the aluminium contact. The EBIC shape reflects the surface texture, however, it is comparatively flattened. 
A line profile across the junction and taken along the dashed gray arrow in Figure \ref{fig:experiment}b is shown in Figure \ref{fig:experiment}c. Including only points below 0.25 to exclude effects of the space charge region, the tails of the profile were fitted to an exponential function on each side yielding a decay length of $83$\,nm and $136$\,nm on the p- and n-side, respectively. Correcting for the approximate 40$^\circ$ inclination of the profile direction with respect to the surface normal, this translates to effective diffusion lengths of
$63$\,nm and $103$\,nm. Obviously, these values are well below those expected in bulk material due to the strong influence of surface recombination which will be discussed later in detail.

In addition to the EBIC experiments, SSRM measurements of another trench were conducted on a cleaved and polished sample edge. The resulting current map is presented in Figure \ref{fig:ssrm}a and line profiles along the labeled directions are plotted in Figure \ref{fig:ssrm}b. In case the mean free path of electrons or holes is significantly shorter than the radius of the scanning tip as well as the length scale on which doping concentrations change considerably, a current minimum is expected at the chemical junction \cite{eyben2002scanning,kalinin2007scanning}. In the presented profiles, minima are found at an approximate depth of 500-550\,nm below the aluminium contact. Since the deviation between repeated SSRM measurements is on the order of 50\,nm -- possibly due to surface charges \cite{prussing2020defect} -- and deviations from the previously mentioned assumptions might lead to a shift of the minimum to the lower-doped side, we consider this value as a rough estimate of the chemical junction position.

\section{Boron Diffusion Simulation}

While sufficiently accurate models exist for the diffusion of boron in silicon, this is neither the case for the growth of boron glasses nor for the diffusion of boron in the glass or for its segregation into silicon. In lack of more detailed knowledge, the main processing steps are modeled as follows:
Following~\cite{1Peter}, the partial pressures of O$_2$ and B$_2$O$_3$
%of approximately 890\,Pa O$_2$ and 100\,Pa B$_2$O$_3$ are established during the deposition step 
under the given conditions in the furnace are calculated yielding the growth of a thin partially liquid boron glass. The doping of the growing glass with boron is simulated using a Dirichlet boundary condition at the interface between the ambience and the glass. After the boron source is switched off, a negligible evaporation of boron into the furnace is assumed during the process steps drive-in, post-oxidation and cool-down. It has to be noted, though, that simulations with diffusion-limited evaporation and an adapted boron concentration at the interface to the ambience during the deposition step lead to very similar boron profiles. 
%The O$_2$ partial pressure during the drive-in and post-oxidation process steps amounts to approximately 1070\,Pa and 67.5\,kPa, respectively, the cool-down step is performed in inert ambience.
Particularly during the post-oxidation step, the boron diffuses deeply into the silicon but segregates also into the growing oxide glass. This results in a retrograde boron profile in the silicon with a maximum concentration approximately 0.12\,\textmu m below the surface.

For the numerical implementation, Sentaurus Process of Synopsys, Version Q-2019.12, was used with advanced calibration and the two-phase segregation model for boron at the silicon-silicon dioxide interface. Within the simple modeling approach taken, the boron concentration at the oxide surface during the deposition step has a main influence on the final sheet concentration of boron in the silicon and thus on its sheet resistance. A second uncertainty results from the enhancement of the boron diffusion during all process steps in oxidizing atmosphere. Established models exist for the dry oxidation of silicon with low dopant concentrations in the oxide. However, the oxide glass growing here is highly doped and partially liquid which should result in considerably less strain in the glass. Following the arguments of~\cite{Cowern2007}, the different stress state should result also in a different injection of self-interstitials and thus oxidation enhancement.

\begin{figure}%
\includegraphics[width=.99\textwidth]{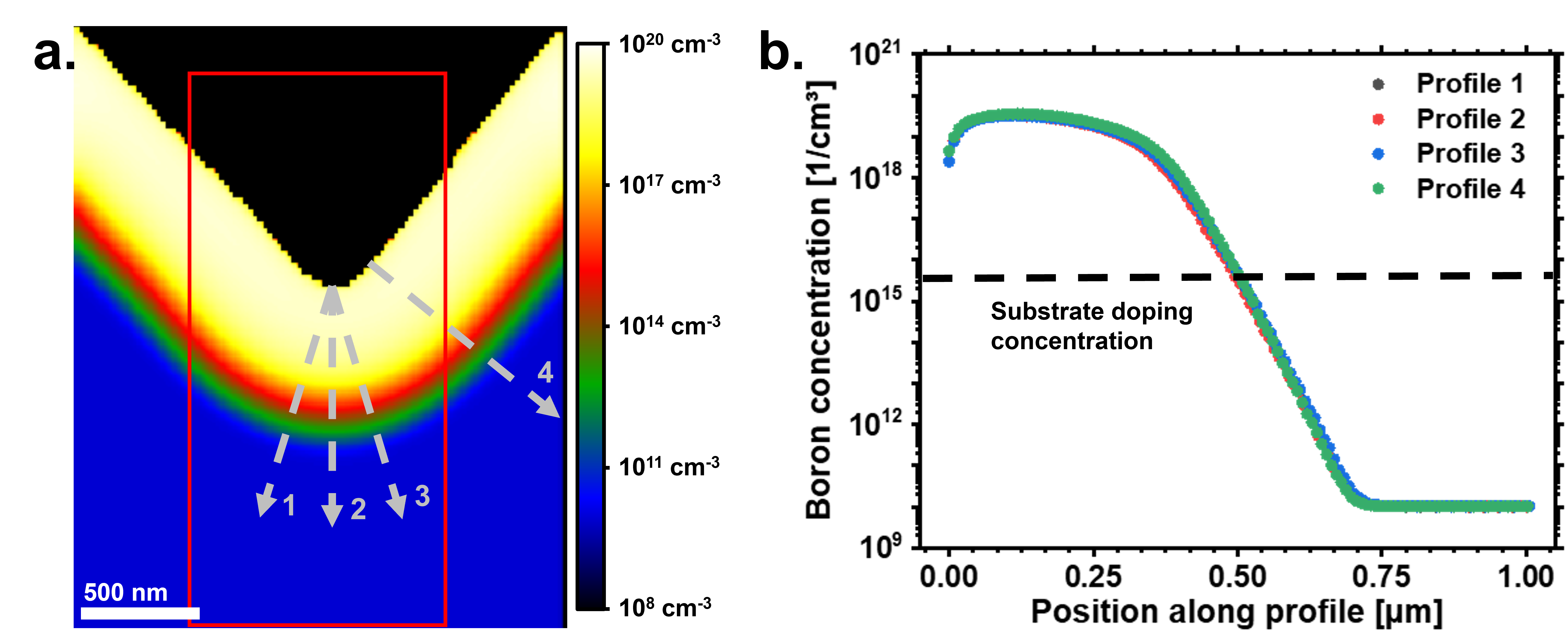}
\caption{Simulated two-dimensional boron concentration underneath the surface texture: The red rectangle marks the region which was scanned in Figure \ref{fig:experiment}a and the model was extrapolated by 500\,nm on the right and left side to avoid edge artifacts in subsequent comparisons to experimental data. The logarithmic colour-code is shown on the right. (b) Line profiles taken along the arrows indicated in (a).}
\label{fig:boron}
\end{figure}

In order to fix the boron concentration at the oxide surface during deposition as well as the degree of oxidation enhancement, two references are used: Firstly, the sheet resistance of 102\,\textOmega/sq obtained by the 4PP measurements on the planar reference sample. Secondly, the position of the chemical junction, i.e., with equal concentrations of ionized acceptors and boron donors of $3.5\times10^{15}$\,cm$^{-3}$, below the tip of the trench in Figure \ref{fig:experiment}a. The latter is assumed to be located 480\,nm below the surface, which is based on the observation that resulting boron distributions used as model input for EBIC simulations leads to positions of the maximal EBIC values consistent with the experiment -- a finding also consistent with the SSRM measurement of a similar trench presented in Figure \ref{fig:ssrm}.
More details about the EBIC simulations are given in the following section and both references are met using an boron surface concentration of $N_\text{B}=1.55\times10^{22}$\,cm$^{-3}$ during the deposition and an oxygen enhanced diffusion (OED) scaling factor of about 0.32 leading to the two-dimensional boron distribution underneath the textured surface shown in Figure \ref{fig:boron}. To avoid boundary effects, the surface topography was linearly extrapolated on the left and right hand side by each 500\,nm in the simulation model and Neumann boundary conditions were assumed for the diffusion equations.

% For the numerical implementation, Synopsis' Sentaurus Process, Version Q-2019.12, was used with the default oxidation models. In order to avoid boundary effects, the surface was linearly extrapolated on the left and right side by each 500\,nm during the simulation.

\section{Numerical EBIC Model}

\begin{table}
    \centering
    \caption{Parameters in equation (\ref{eq:poisson})-(\ref{eq:rec}) used for the FEM simulations.}
    \begin{tabular}{cccccccc}
    \hline\hline
    Parameter&$\epsilon_r$&$\mu_\text{e}$&$\mu_\text{h}$&$T$&$n_i$&$\tau_\text{h}$&$\tau_\text{e}$\\\hline
    Value&11.7&1100\,cm$^2$V$^{-1}$s$^{-1}$&200\,cm$^2$V$^{-1}$s$^{-1}$&300\,K&9.802$\times10^{9}$cm$^{-3}$&1.988$\times10^{-11}$\,s&7.130$\times10^{-13}$\,s\\\hline\hline
    \end{tabular}
    \label{tab:parameters}
\end{table}

\begin{figure}
\includegraphics[width=.99\textwidth]{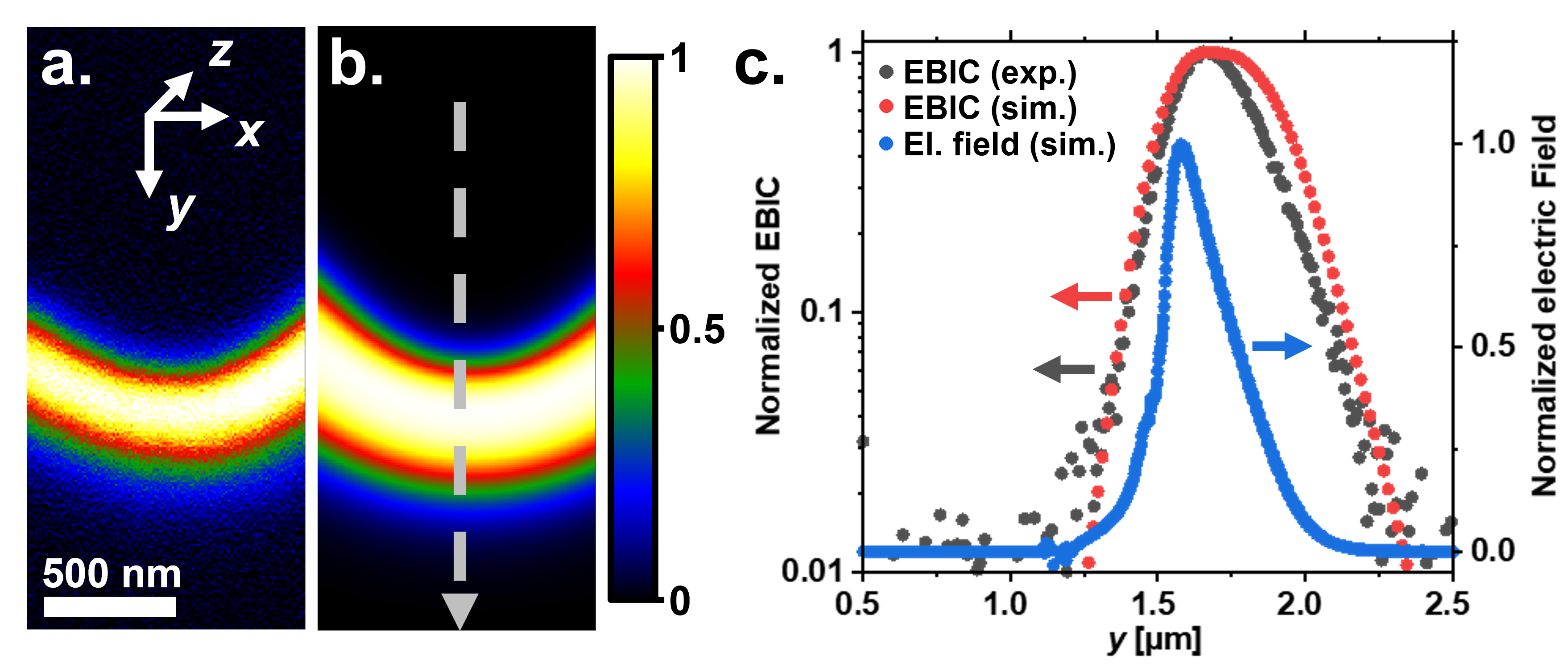}
\caption{(a) experimental and (b) simulated EBIC map normalized to their maxima. (c) Vertical profiles of the experimental and simulated EBIC (logarithmic scale to the left) as well as the normalized magnitude of the electric field obtained in the FEM simulation (linear scale to the right) taken along the dashed gray line in (b).}
\label{fig:comparison}
\end{figure}

To simulate the drift diffusion behaviour of excess charge carriers, a 2D model (omitting the electron beam propagation direction, i.e. $z$) with three variables is used: The electrostatic potential $\phi$ as well as the electron and hole concentrations $n$ and $p$, which are linked via the van-Roosbroeck set of differential equations including Poisson's equation for the electrostatic potential and the stationary continuity equations for electrons and holes \cite{van1950theory}: 
\begin{align}
    -\vec{\nabla}\left(\epsilon_0\epsilon_\text{r}\vec\nabla\phi\right)=e\left(N_\text{d}-N_\text{B}+p-n\right)\label{eq:poisson}\,,\\
    0=\frac{\text{d}n}{\text{d}t}=-\vec\nabla\left(\mu_\text{e}n\vec\nabla\phi-\frac{k_\text{B}T}{e}\mu_\text{e}\vec\nabla n\right)+g-r\label{eq:n}\,,\\
    0=\frac{\text{d}p}{\text{d}t}=-\vec\nabla\left(-\mu_\text{h}p\vec\nabla\phi-\frac{k_\text{B}T}{e}\mu_\text{h}\vec\nabla p\right)+g-r\label{eq:p}\,.
\end{align}
Here, $N_\text{d}$ is the ionized donor concentration and $N_\text{B}$ the boron concentration presented in Figure \ref{fig:boron}, which is considered as the ionized acceptor concentration. Furthermore, the relative permittivity $\epsilon_\text{r}$, the electron and hole mobilities $\mu_\text{e}$ and $\mu_\text{h}$, the intrinsic charge carrier concentration $n_\text{i}$ as well as the excess charge carrier generation and recombination rates $g$ and $r$ are used. The two parenthesized terms on the right side of equation (\ref{eq:n}) and (\ref{eq:p}) express drift and diffusion of charge carriers and the Einstein relation has been used to express the latter with mobilities instead of diffusivities. 
The values of the constant parameters used in the simulations are summarized in Table \ref{tab:parameters}. 
The concentration of ionized donors is held constant at 3.5$\times10^{15}$\,cm$^{-3}$, which corresponds to the wafer resistivity of 1.4\,\textOmega cm.

The generation rate $g$ due to the electron beam is modeled as a Gaussian, where the mean value represents the position of the electron beam. The standard deviation was set to 10\,nm to guarantee an extension similar to the mesh size and the amplitude to 8.87$\times10^{25}$\,cm$^{-3}$s$^{-1}$.
The recombination rate $r$ is expressed by the following term which is motivated by a single Shockley-Read-Hall recombination centre \cite{Black2016} (but omits re-emission from the trap state):
\begin{align}
    r=\left(np-n_\text{i}^2\right)/\left(\tau_\text{h}n+\tau_\text{e}p\right)\,.\label{eq:rec}
\end{align}
This definition aims to describe an effective lifetime including both bulk recombination as well as those caused by defect states due to the surfaces. In the space charge region where electron and hole concentrations are similar, the recombination rate depends on both parameters $\tau_\text{h}$ and $\tau_\text{e}$.
In the neutral regions, however, it is dominated by $\tau_\text{h}$ on the n-side and by $\tau_\text{e}$ on the p-side. Thus, the two parameters can be set independently to resemble the decay lengths found experimentally in Figure \ref{fig:experiment}c, which is achieved for the values 
given in in Table \ref{tab:parameters}. 
Re-emission from the defect level is excluded as it demands for a third input parameter, i.e. the unknown trap energy level. The recombination parameters are held constant in the model as the standard deviation of the thickness (8\,nm) is much lower than the average thickness (312\,nm) in the scanned area and no significant change in the interplay of surface and bulk recombination is expected.
%Similarities to existing approaches in the literature as well as limitations of this model will be discussed later in detail.

The boundary conditions at the EBIC contacts (top and bottom surface) are chosen to be Dirichlet conditions with $\phi=0$, $n=n_i^2/N_\text{B}$, and $p=N_\text{B}$ at the p-side and $\phi=k_\text{B}T/e\log{\left(N_\text{d}N_\text{B}/n_i^2\right)}$, $n=N_\text{d}$, and $p=n_i^2/N_\text{d}$ at the n-side.
The EBIC current is evaluated at the n-side contact.
The boundary conditions in $x$ direction are chosen to be zero flux Neumann conditions, i.e. the $x$ component of the gradient of $\phi$, $n$, and $p$ are set to zero. 

For the numerical FEM implementation,  COMSOL Multiphysics, Version 5.4., was used. In order to avoid boundary effects, the experimentally scanned region was (in accordance with the boron diffusion simulation shown in Figure \ref{fig:boron}) extrapolated by each 500\,nm on the left and right side.

\section{Comparison of Experiment and Simulation}

While the previous three sections gave detailed descriptions of the experimental results as well as the boron diffusion and EBIC simulations, this section focuses on comparing the corresponding findings and illuminating the connections between them. In fact, the three parts build a self-consistent framework since the OED scaling factor in the boron diffusion simulation was fixed such that the maximum of a subsequent EBIC simulation coincides with the experimentally obtained position.
The resulting experimental and simulated EBIC maps normalized to their maxima are presented in Figure \ref{fig:comparison}a and \ref{fig:comparison}b showing a good agreement in the signal shape. 
In addition to the two-dimensional maps, both EBIC signals as well as the electric field magnitude along the profile indicated by the dashed gray arrow in Figure \ref{fig:comparison}b are given in Figure \ref{fig:comparison}c.
As enforced in the chosen self-consistent framework, the EBIC maxima occur at the same position. The location of the strongest electric field, however, which coincides with the chemical junction satisfying $N_\text{B}=N_\text{d}=3.5\times10^{15}$\,cm$^{-3}$, is shifted relatively by approximately 80\,nm.
Furthermore, the profile tails of the experimental and simulated EBIC signals exhibit consistent decay lengths. This finding indicates that a uniform choice of the recombination parameters -- fixed on the left boundary of the experimentally investigated region -- successfully describes the interplay of bulk and surface recombination in the entire scanning range outside the space charge region.
In contrast, deviations between experiment and simulation are visible inside the latter, particularly in the n-type region to the right of the junction. Possible origins for this discrepancy are (i) surface charges leading to an alteration of the electric field not included in the simulation model, (ii) details in the recombination process that are not captured in the simplified model given by Equation (\ref{eq:rec}), and (iii) heterogeneity along the electron $z$-direction, which is omitted in the two-dimensional simulations.
Since the investigated trench is defined by two intersecting (111) planes and expected to be aligned with the electron beam, we consider (iii) as unlikely.
Surface charges, however, serve as a probable explanation since they
influence p- and and n-type regions differently depending on the Fermi level of related defects \cite{Haney2016} and the increased deviation between experiment and simulation on the n-side are consistent with surface acceptor states, e.g., due to gallium implantation during FIB preparation.
Nevertheless, (ii) cannot be ruled out completely and more sophisticated recombination terms, e.g., including ensembles of defect states with different energy levels, might lead to more accurate modeling of the EBIC signal inside the space charge region.

\begin{figure}
\centering
\includegraphics[width=.49\textwidth]{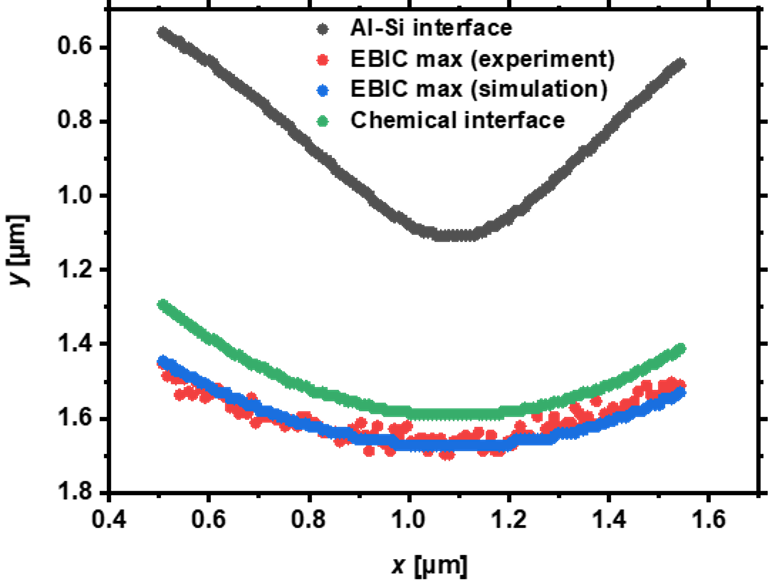}
\caption{Relevant contour lines: Al-Si interface position (gray) extracted from Figure \ref{fig:experiment}a as local maximum after Sobel filtering. Column-wise maximal values of the experimental (red) and simulated (blue) EBIC maps presented in Figure \ref{fig:comparison}a and \ref{fig:comparison}b. In addition, the position of the chemical interface satisfying $N_\text{B}=N_\text{d}=3.5\times10^{15}$\,cm$^{-3}$ and obtained from the simulated boron concentration in Figure \ref{fig:boron} is shown (green).}
\label{fig:contours}
\end{figure}

To enable a comparison beyond individual line profiles, several contour lines in the $xy$-plane are plotted in Figure \ref{fig:contours} including the aluminium-silicon interface (gray), the experimental (red) and simulated (blue) column-wise EBIC maxima, as well as the chemical interface (green) resulting from the boron diffusion simulation. The overall agreement of the EBIC maxima positions underlines the self-consistency of the results not only in the line profiles used to determine relevant parameters, but across the entire investigated area.
In addition, the plot reveals the shape similarity of the chemical junction and the EBIC maxima and highlights the aforementioned shift of approximately 80\,nm between them.
We emphasize at this point that such shifts are expected even for symmetrically doped junctions due to different mobilities of electrons and holes.

\section{Conclusion}

Electron beam induced current (EBIC) measurements conducted in a scanning transmission electron microscope (STEM) offer great potential to study excess charge carrier dynamics, electric fields, and to trace dopant distributions in micro- and nanoscale functional devices. In this paper, a textured silicon solar cell with a boron diffused emitter was investigated experimentally and modeled in a self-consistent framework including boron diffusion and EBIC simulations. 

A key achievement is the demonstration that two-dimensional EBIC maps can be simulated and compared to experimental results of laterally heterogeneous samples. This was accomplished by using an effective recombination model -- as recently suggested in \cite{meyer2026quantitativemodelsexcesscarrier} -- to describe the effect of surfaces in the electron transparent sample, i.e., by reducing the three-dimensional experimental geometry to an effectively two-dimensional.
The approach leads to an excellent agreement between simulation and experiment in neutral semiconductor regions, which is consistent with an effective diffusion length for homogeneous materials in the presence of surface recombination.
Inside the space charge region, simulated EBIC values are higher than in the experiment suggesting an electric field dependence of the used effective recombination parameters. Experimentally, this could be tested in future studies by combining STEM-EBIC with electron holography \cite{anada2020direct,lindner2024reconstruction} or four-dimensional STEM \cite{wartelle2025sub,flathmann2025sequential}. 
Importantly, the low doping concentration of the substrate used in this study shows that once a model for the electric field dependence is established, STEM-EBIC might serve as a highly sensitive method to measure electric fields beyond the sensitivity limit of the mentioned complementary techniques.
Lastly, besides the progress in modeling excess carrier recombination and STEM-EBIC signals, the presented work clearly demonstrates how valuable feedback about dopant distributions can be obtained in heterogeneous geometries and on the nanometer scale.

In summary, we have shown that modeling of two-dimensional STEM-EBIC maps is feasible, which opens up unprecedented possibilities to observe excess carrier dynamics, weak electric fields and dopant distributions in nanoscale functional devices.

\medskip
\noindent\textbf{Acknowledgements}

We thank Jan Verhoeven for valuable discussions about the used electronic setup as well as Matthias Hahn, Volker Radisch, and Thomas Lehmann for technical support. 
T.M, C.F., and M.S. acknowledge funding by the Deutsche Forschungsgemeinschaft (DFG, German Research Foundation) -- 217133147/SFB 1073, project B02. 
V.T. acknowledges funding via the PhD Scholarship Programme of the  German Federal Environmental Foundation (DBU).
J.S. acknowledges funding by German State of Lower Saxony.
M.S. acknowledges funding by the Deutsche Forschungsgemeinschaft (DFG, German Research Foundation) -- project 429413061 (SE560/8-1).
The use of equipment of the “Collaborative Laboratory and User Facility for Electron Microscopy” (CLUE, Göttingen) is gratefully acknowledged.

\medskip
\noindent\textbf{Author Contributions}

T.M.:
conceptualization; 
data curation;
formal analysis;
investigation;
methodology;
software;
validation;
visualization;
writing—original draft;
writing—review \& editing;
supervision.
D.A.E.:
conceptualization; 
formal analysis;
investigation;
methodology;
software;
validation;
visualization;
writing—review \& editing;
P.P.:
formal analysis;
investigation;
validation;
visualization;
writing—original draft;
writing—review \& editing;
K.L.S.D.:
formal analysis;
investigation;
methodology;
validation;
visualization;
writing—review \& editing;
C.F.:
investigation;
writing—review \& editing;
supervision.
V.T.:
formal analysis;
investigation;
writing—review \& editing;
funding acquisition;
T.B.:
investigation;
writing—review \& editing.
H.B.:
writing—review \& editing;
resources;
funding acquisition;
supervision.
J.S.:
writing—review \& editing;
resources;
funding acquisition;
supervision.
M.S.:
conceptualization; 
writing—review \& editing;
resources;
funding acquisition;
supervision;
project administration.

\medskip
\noindent\textbf{Data Availability Statement}

The data that support the findings of this study are openly available in \cite{publication_data} and additional raw data are available from the corresponding author upon reasonable request.

\bibliographystyle{MSP}
\bibliography{bibliography}

%\textbf{References}\\ % uncomment and insert .bbl file content here before submssion

% Figures/tables and captions
% Permission statements are required for all figures reproduced or adapted from previously published articles/sources. Please also ensure that all necessary permissions to reproduce images have been received
% Please remove these statements for original figures

\begin{figure}
\textbf{Table of Contents}\\
\medskip
  \includegraphics[width=\linewidth]{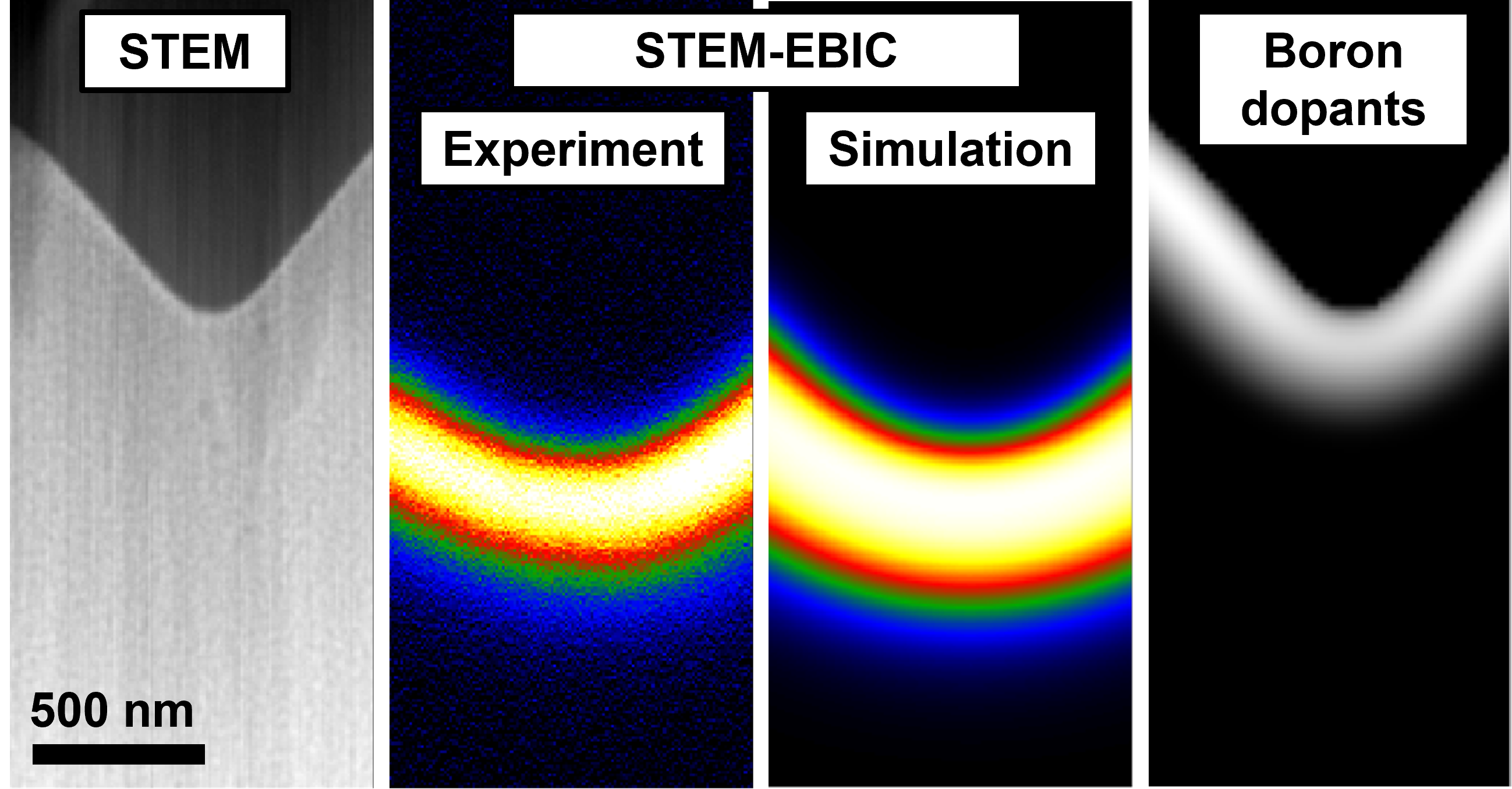}
  \medskip
  \caption*{
Self-consistent modeling of electron beam induced current (EBIC) signals and boron distribution in a textured silicon solar cell. Using an effective lifetime model enables two-dimensional EBIC simulations and direct comparison two experimental data allowing refinement of the boron distribution.
  }
\end{figure}

\end{document}